\documentclass[journal,comsoc]{IEEEtran} 

\usepackage[T1]{fontenc}

\usepackage{lipsum}  

\usepackage{lineno}
\usepackage{multirow}
\usepackage{booktabs}
\usepackage{tabulary} 
\usepackage{dingbat}
\usepackage{amsmath}

\usepackage{color, soul}
\usepackage[dvipsnames,table]{xcolor}
\usepackage{hhline}
\usepackage[skip=1pt]{subcaption}
\usepackage{caption}
\usepackage[normalem]{ulem} 
\usepackage{enumitem} 
\usepackage{tikz}
\usetikzlibrary{positioning}
\usetikzlibrary{shapes,arrows}
\usetikzlibrary{decorations.pathreplacing}
\usepackage{paralist} 

\usepackage{cite}

\newcommand{\specialcell}[2][c]{%
  \begin{tabular}[#1]{@{}l@{}}#2\end{tabular}}

\ifCLASSINFOpdf
\else
\fi

\usepackage{amsmath}

\usepackage[cmintegrals]{newtxmath}

\begin{document}

\title{Selfish Attacks in Two-hop IEEE 802.11 Relay Networks: Impact and Countermeasures}
\author{Szymon~Szott~
        and~Jerzy~Konorski\vspace{-0.5cm}
\thanks{S. Szott is with the Faculty of Computer Science, Electronics and Telecommunications,
AGH University (e-mail: szott@kt.agh.edu.pl). 
}
\thanks{J. Konorski is with the 
Faculty of Electronics, Telecommunications and Informatics,  
Gdansk University of Technology (e-mail: jekon@eti.pg.gda.pl).}
}

\maketitle

\begin{abstract}
In IEEE 802.11 networks, selfish stations can pursue a
better quality of service (QoS) through selfish MAC-layer attacks.
Such attacks are easy to perform, secure routing protocols do not prevent them, and their detection may be complex.
Two-hop relay topologies allow a new angle of attack: a selfish relay can tamper with either source traffic, transit traffic, or both.
We consider the applicability of selfish attacks and their variants in the two-hop relay topology, quantify their impact, and study defense measures.
\end{abstract}


\IEEEpeerreviewmaketitle

\section{Introduction}

\begin{figure}[!t]
\centering
\includegraphics[width=0.8\columnwidth]{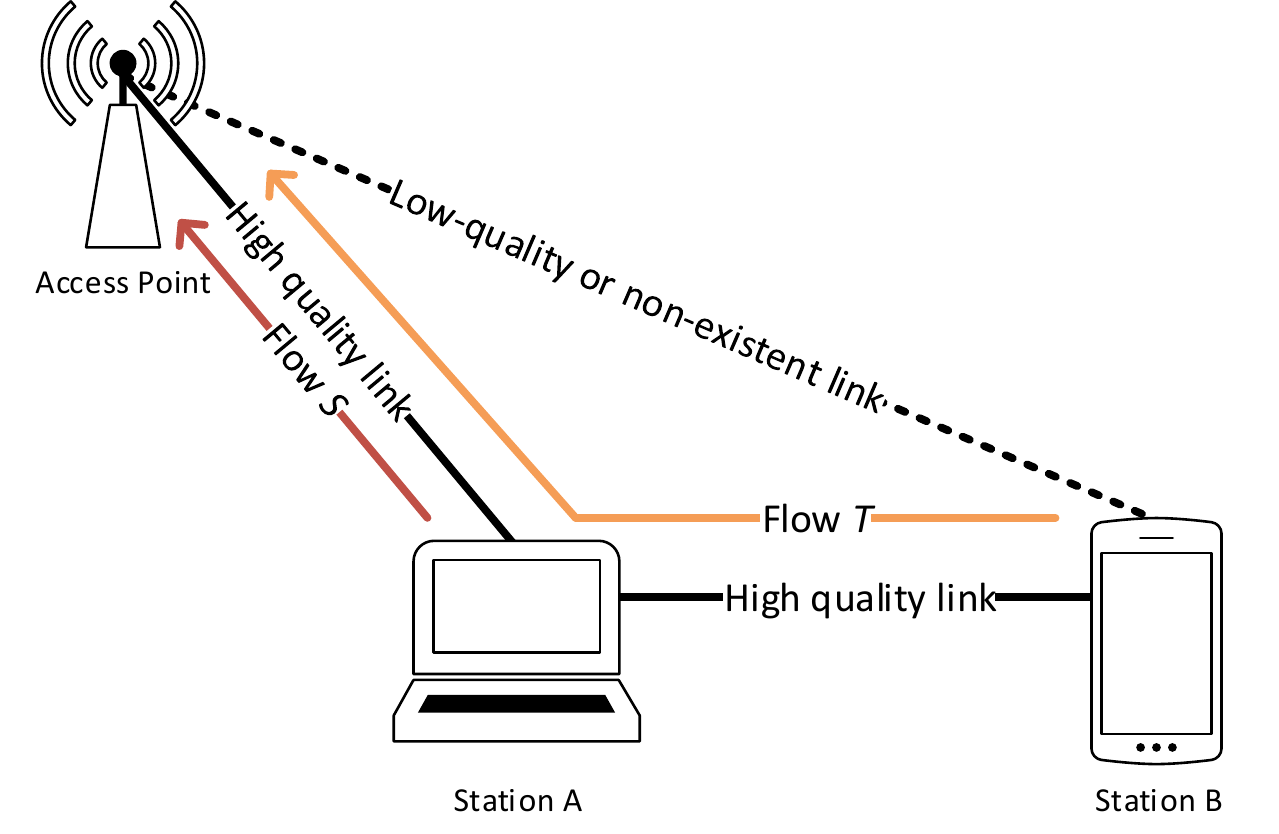}
\caption{Conceptual setting for selfish attacks in a two-hop 
relay network (all nodes with one wireless interface communicating on a single channel, neighboring nodes within carrier sensing range). 
Station A, as the relay, can increase its QoS perception by executing a selfish attack: upgrading source traffic (flow $S$), downgrading transit traffic (flow $T$), or both.}
\label{fig:topology}
\end{figure}

\IEEEPARstart{T}{he} coverage of IEEE 802.11 networks can be extended if stations connected to an access point (AP) act as relays, i.e., they share their connection with other, neighboring stations, who either cannot reach the AP themselves or have a poor direct connection to it. 
This approach creates a two-hop relay network (Fig. \ref{fig:topology}) which is known to have many advantages in terms of network coverage and  performance \cite{Garcia-Saavedra2015}. 

A two-hop relay network requires cooperation from the relay station. If it declares cooperation,
the relay may gain a privileged status at the AP, e.g., enjoying better terms of network access.  
In the following, we consider the specific case that the AP is a public hotspot and Internet access is granted free-of-charge provided that all one-hop connected stations act as relays to improve network performance\footnote{Technically, this requires that the relay act as a router, using either the path selection protocol of 802.11s or any ad hoc routing protocol.}.
In this setting, the relay may now be motivated to launch \emph{selfish attacks} resulting in preferential treatment for its own (source) traffic over relayed (transit) traffic, hence to achieve an undue (and possibly undetected) increase of the quality of service (QoS) \cite{Konorski2014}. 

Selfish attacks can be launched in two ways. First, packet scheduling in the forwarding path can be biased in favor of source traffic. Second, source packets can be unduly prioritized at the MAC-layer. Since it is the MAC mechanisms that ultimately decide the order and delays of medium acquisition by successive packets (Section~\ref {sec:qos}), in what follows we focus on MAC-layer attacks 
defined in Section~\ref{sec:Selfish}.

Selfish MAC-layer attacks pose a serious threat to IEEE 802.11 networks: they are easy to perform, secure routing protocols do not prevent them, and their detection may be complex \cite{Konorski2014}.
These attacks have been studied in single-hop networks \cite{Konorski2014}.
However, the two-hop relay topology adds another dimension to the problem, in that the relay can tamper with either source traffic, or transit traffic, or both (Fig. \ref{fig:topology}).
We consider the applicability of these attacks and their variants in the two-hop relay topology (Section \ref{sec:linear}), quantify their impact  (Section \ref{sec:impact}), and study defense measures (Section \ref{sec:countermeasures}).
In our view, the presented attack scenario and proposed defence measures can serve as a useful analytical framework in any 2-hop configuration with a relay, in particular envisaged for 5G femtocells, regardless of the routing protocol and TRA method used.

\section{QoS Provisioning in IEEE 802.11 Networks}
\label{sec:qos}

QoS provisioning in IEEE 802.11 is achieved through the enhanced distributed channel access (EDCA) function,
where higher-layer traffic classes are mapped to queues of one of four access categories (ACs).
The ACs, in order of decreasing priority, are
voice (VO), video (VI), best effort (BE), and background (BK).
The configuration of each AC 
provides statistical prioritization with respect to channel access and duration \cite{Konorski2014}.

Traffic classification into ACs is based on the Differentiated Services Code Point (DSCP) set in a packet's IP header: in the \texttt{Type of Service} (ToS) field in IPv4 or \texttt{Traffic Class} (TC) field in IPv6. 
DSCP values can be configured according to higher-layer policies, using network-layer packet mangling software (such as Linux \texttt{iptables})
for all packets belonging to a given flow.

\section{Selfish Attacks in IEEE 802.11 Networks}
\label{sec:Selfish}
The QoS provisioning model of IEEE 802.11 networks enables two types of selfish attacks: 
backoff attacks (BOAs) and traffic remapping attacks (TRAs).
Both can be executed either as \emph{source traffic upgrading} or \emph{transit traffic downgrading} (Fig.~\ref{fig:topology}), which we denote ($^+$) and ($^-$), respectively. 
We consider only BOAs and TRAs because they are relatively easy to execute, in contrast to, e.g., forging frame headers \cite{Konorski2014}.

\subsection{Backoff Attacks}
\label{sec:boa_disc}
BOAs belong to the class of MAC parameter manipulation attacks -- out of the available medium access parameters 
the contention window (CW) has proved to be the easiest to manipulate \cite{cagalj2005selfish,Patras2016}.
The CW governs the backoff mechanism, wherein each station \emph{backs off} before accessing the
channel by waiting a random number of time slots.
An attacker may attempt to influence this random behavior so as to improve its QoS, either by decreasing CW for the AC of source traffic (BOA$^+$) or increasing it for transit traffic (BOA$^-$).

BOAs have 
two key advantages. First, modifying MAC parameters is often
available as part a user's configuration interface. 
Second, detecting BOAs is challenging due to both the
randomness inherent to the backoff function as well as the practical difficulties in performing precise time measurements. 

The BOA$^+$ attack has mostly been studied in a single-hop infrastructure-based 802.11s setting and shown to effectively promote the attacker's traffic \cite{cagalj2005selfish}. 
In a two-hop topology, a BOA$^-$ can additionally be applied to transit traffic to discriminate it in favor of source traffic. 
Research is required to quantify the impact of BOAs in two-hop relay settings.

\subsection{Traffic Remapping Attacks}
\label{sec:tra_disc}

TRAs consist in claiming a different medium access priority through false DSCP settings so that traffic can be mapped onto a different AC. 
TRAs are simpler to perform than BOAs: a user application can access packet mangling software to change the current DSCP of any packet. 

Upgrading TRAs (TRA$^+$) have been studied in single-hop ad hoc networks \cite{Konorski2014}, where a distributed
discouragement scheme, based on the threat of detection and punishment, allows TRAs only if they are harmless to honest stations; otherwise it induces selfish stations to learn that a long-sustained TRA is counterproductive. For
relay networks, TRAs can be performed also on transit traffic to lower its
priority (TRA$^-$). 
Such a possibility has thus far only been studied in a multi-hop ad hoc network \cite{Konorski2017}, necessitating further research for the specific two-hop relay setting.

\section{Two-Hop Relay Topology}
\label{sec:linear}

We make the following assumptions for analyzing the network in Fig.~\ref{fig:topology}.
Station B and the AP are out of communication range. The placement of A in the topology allows the execution of any one of the previously discussed attacks. 
For ease of presentation we reduce the configuration space assuming that only two ACs are used: VO and BE. 
We consider the traffic transmitted in these ACs as a generic representation of high and low priority traffic, respectively.
The interesting case for analysis is when, at A, the transit flow $T$ is VO and the source flow $S$ is BE. 
We evaluate the attack performance under saturation. 
We consider TCP because saturated UDP traffic causes a fairness problem -- if a station's queue is full of source traffic then it drops all transit traffic. 
This is not the case for the self-regulating TCP. 

\subsection{Uplink and Downlink Scenarios}

In the \textit{uplink scenario}, there are two saturated TCP data flows terminating at the AP upon which A can execute the attacks: $S$, referred to as the source flow, and $T$, referred to as the transit flow. 
A's goal is to improve its uplink throughput (i.e., that of $S$). There are also traffic flows carrying TCP ACK segments in the reverse direction: $S'$ and $T'$ (omitted from Fig.~\ref{fig:topology} for clarity).
These two flows are important: although they have a low rate, they impact the time spacing of the TCP data and thus end-to-end throughput of flows $S$ and $T$.

A \textit{downlink scenario} can also be considered, where the saturated TCP data flows are $T$ and $S'$. Note that again A can directly influence $T$ and $S$, the latter now consisting of TCP ACKs for flow $S'$. A's goal is now to improve its downlink throughput (i.e., that of $S'$). Whether MAC-layer attacks can have any impact on downlink throughput for two-hop relay settings is an open question that we want to address. 

\subsection{Attack Strategies}
\label{sec:variants}

For each attack strategy it is helpful to specify which AC queues are used at the attacker, how they are configured, and which traffic is sent using which AC (Table~\ref{tab:comp1}). 
For BOAs, we assume that AC queues are configured with valid EDCA parameters, e.g., in BOA$^+$ the BE queue is configured with VO parameters.
Note that BOA$^+$ and TRA$^+$ differ in their effects even though they share the same configuration for $S$ and $T$; similarly for BOA$^-$ and TRA$^-$. This is because under BOA$^+$ and BOA$^-$ both AC queues are used, whereas TRA$^+$ and TRA$^-$ merge source and transit traffic into a single AC queue, causing less inter-queue contention at the MAC layer. They also modify the QoS designation of each packet, which impacts its end-to-end as well as TCP ACK transmission priority, whereas BOA$^+$ and BOA$^-$ have local impact only.

\begin{table}[t]
\centering
\caption{Comparison of attacker's MAC-layer configuration for the BOAs and TRAs in the setting of Fig.~\ref{fig:topology}. 
The last two columns indicate, respectively, which EDCA queue is used and what is its configuration. The latter denotes priority in medium access during frame transmission, the former -- the QoS designation of the frames.} 
\label{tab:comp1}
\begin{tabular}{@{}llll@{}}
\toprule
\specialcell{Attack strategy} & \specialcell{Flow,\\intrinsic AC} & \specialcell{AC queue\\used} & \specialcell{Queue\\configured as}
      \\ \midrule
\multirow{2}{*}{\specialcell{None\\(honest behavior)}} 			& T, VO  & VO            & VO           \\
                      											& S, BE  & BE            & BE \\ \cmidrule(l){2-4} 
\multirow{2}{*}{BOA$^+$} 										& T, VO  & VO            & VO  \\
                     											& S, BE  & BE           & VO               \\  \cmidrule(l){2-4} 
\multirow{2}{*}{BOA$^-$} 										& T, VO  & VO            & BE  \\
                      											& S, BE  & BE           & BE               \\  \cmidrule(l){2-4} 
\multirow{2}{*}{TRA$^+$} 										& T, VO  & VO            & VO  \\
                      											& S, BE  & VO            & VO   \\  \cmidrule(l){2-4} 
\multirow{2}{*}{TRA$^-$} 										& T, VO  & BE 			  & BE\\
                      											& S, BE  & BE           & BE                \\  \cmidrule(l){2-4} 
\multirow{2}{*}{\specialcell{2xTRA\\(TRA$^+$ and TRA$^-$)}} 	& T, VO  & BE           & BE  \\
                      											& S, BE  & VO           & VO     \\  
\bottomrule
\end{tabular}
\end{table}

BOAs and TRAs can also be combined into more sophisticated attack strategies. 
In particular, BOA$^+$ and BOA$^-$ (2xBOA) can be combined to produce a priority switch between the source and transit traffic flows ($S$ and $T$). A similar priority switch is produced by a combination of TRA$^+$ and TRA$^-$ (2xTRA).

In combinations of BOA and TRA, the TRA component assigns both flows to the same AC queue, for which the BOA component increases or decreases CW and other EDCA parameters. Thus the provided QoS is the same for both flows and only depends on the effectiveness of the underlying contention resolution. Hence, combinations of BOA and TRA come under performance optimization rather than network security, and as such are outside our scope. Henceforth we consider only 2xBOA and 2xTRA.

\section{Attack Impact Analysis}
\label{sec:impact}

To study the impact of the considered attacks we used the ns-2.28 simulator. 
A multi-hop IEEE 802.11b\footnote{Simulations were also performed for newer IEEE 802.11 physical layers with the same qualitative results.} network 
served the three stations in Fig.~\ref{fig:topology}.
Important simulation settings include: enabling Request to Send/Clear to Send, setting transmission and interference range at one and two hops, respectively, using ad hoc on-demand distance vector routing, generating 1000-byte data packets, and using transmission queues of 50 packets.
Simulations lasted 90 s (with a 15 s warm-up). Each data point is the mean of 20 experiment runs.

The provided QoS is defined based on a flow's intrinsic AC: as packet delay for VO traffic, i.e., $T$ (which according to ITU-T recommendations should not exceed 100 ms), and as achieved throughput for BE traffic, i.e., $S$ or $S'$ in the uplink or downlink scenario, respectively. The selfish relay A attempts to maximize the throughput of $S$ or $S'$ at the cost of the victim flow $T$ while maintaining a low risk of detection. We have simulated all attack strategies available to A (Table \ref{tab:comp1}). We have omitted from the figures attack strategies combining BOA and TRA (indicated above as out of scope), as well as 2xBOA, which was found only a marginal improvement over stand-alone BOA$^-$ (the contribution of BOA$^+$ turned out to be negligible).

\subsection{Uplink Scenario}
\label{sec:uplink}

\begin{figure*}[tbp]
\centering
\begin{subfigure}[]{0.32\textwidth}
\includegraphics[width=\textwidth]{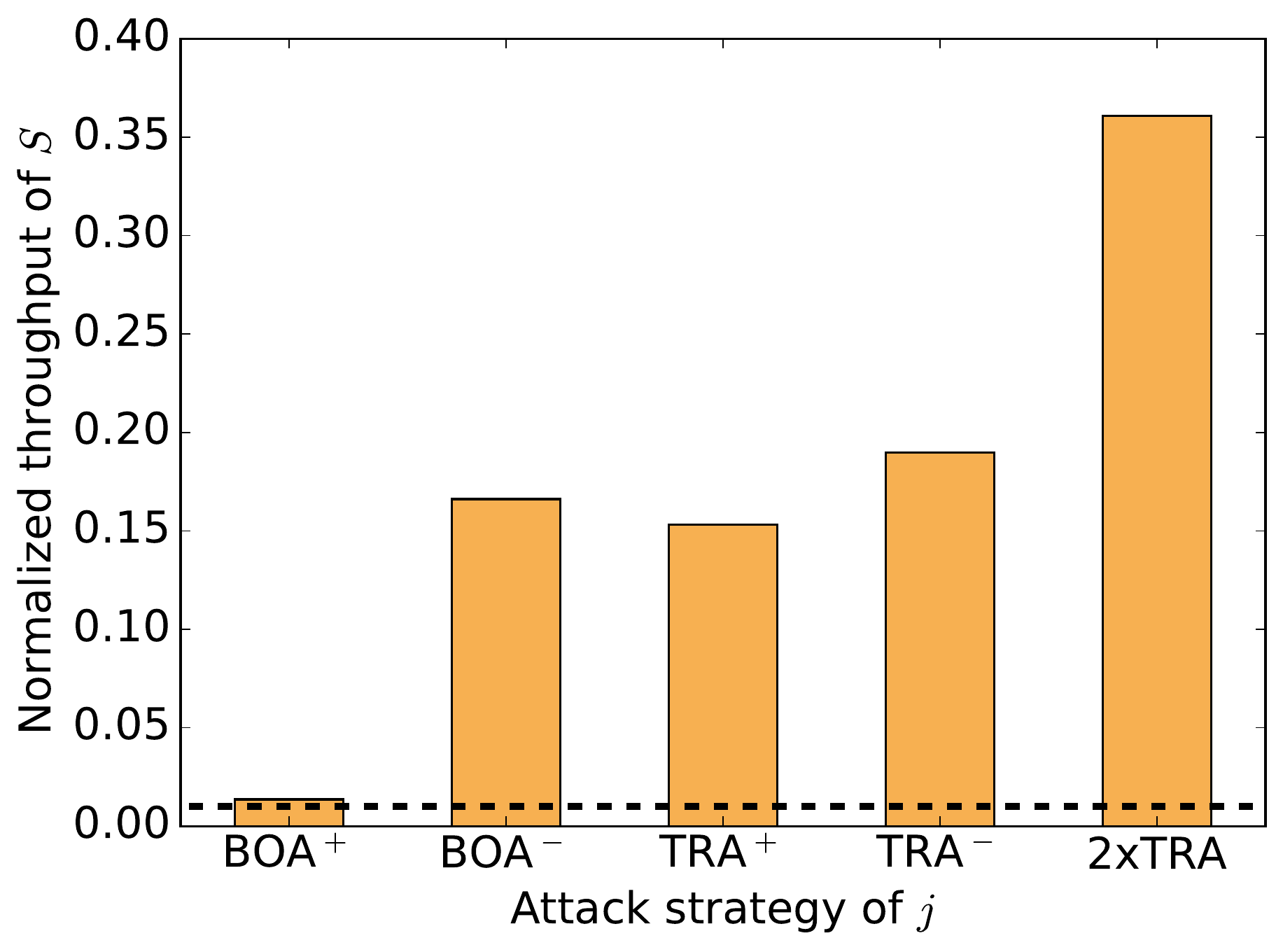}
\caption{}
\label{fig:singlehop-sat-ul-j}
\end{subfigure}
\begin{subfigure}[]{0.32\textwidth}
\includegraphics[width=\textwidth]{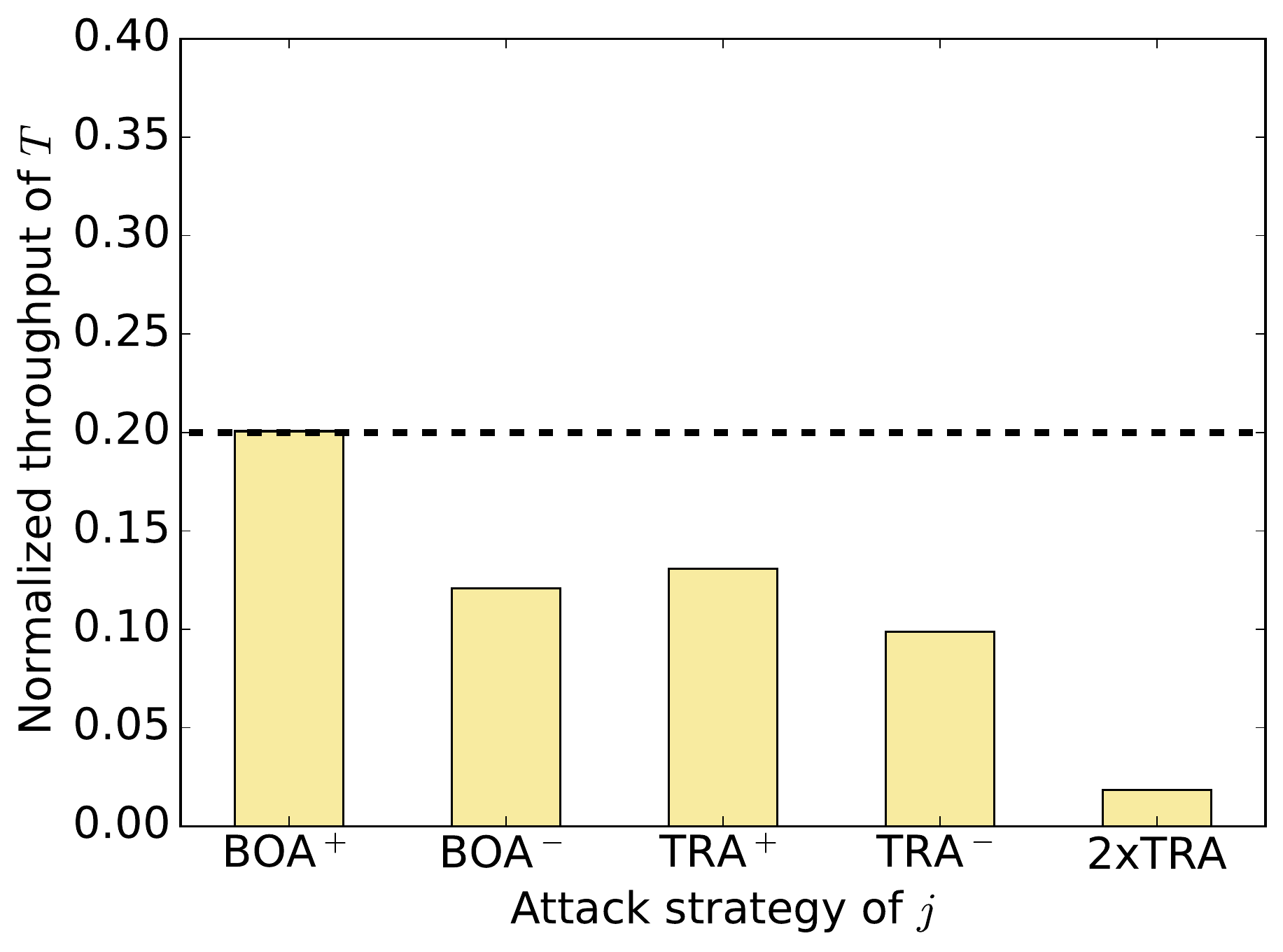}
\caption{}
\label{fig:singlehop-sat-ul-i}
\end{subfigure}
\begin{subfigure}[]{0.32\textwidth}
\includegraphics[width=\textwidth]{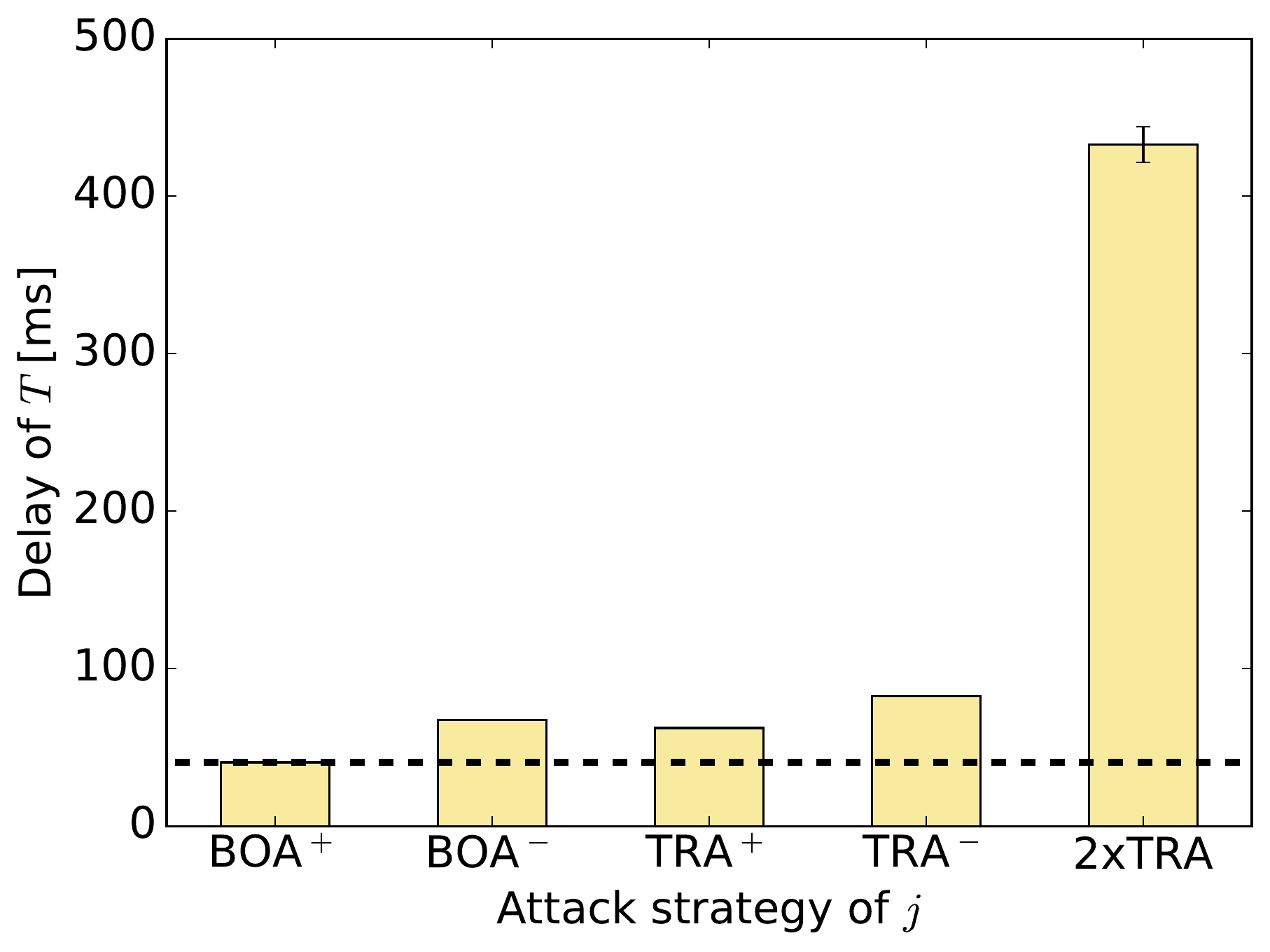}
\caption{}
\label{fig:singlehop-sat-ul-del}
\end{subfigure}
\begin{subfigure}[]{0.32\textwidth}
\includegraphics[width=\textwidth]{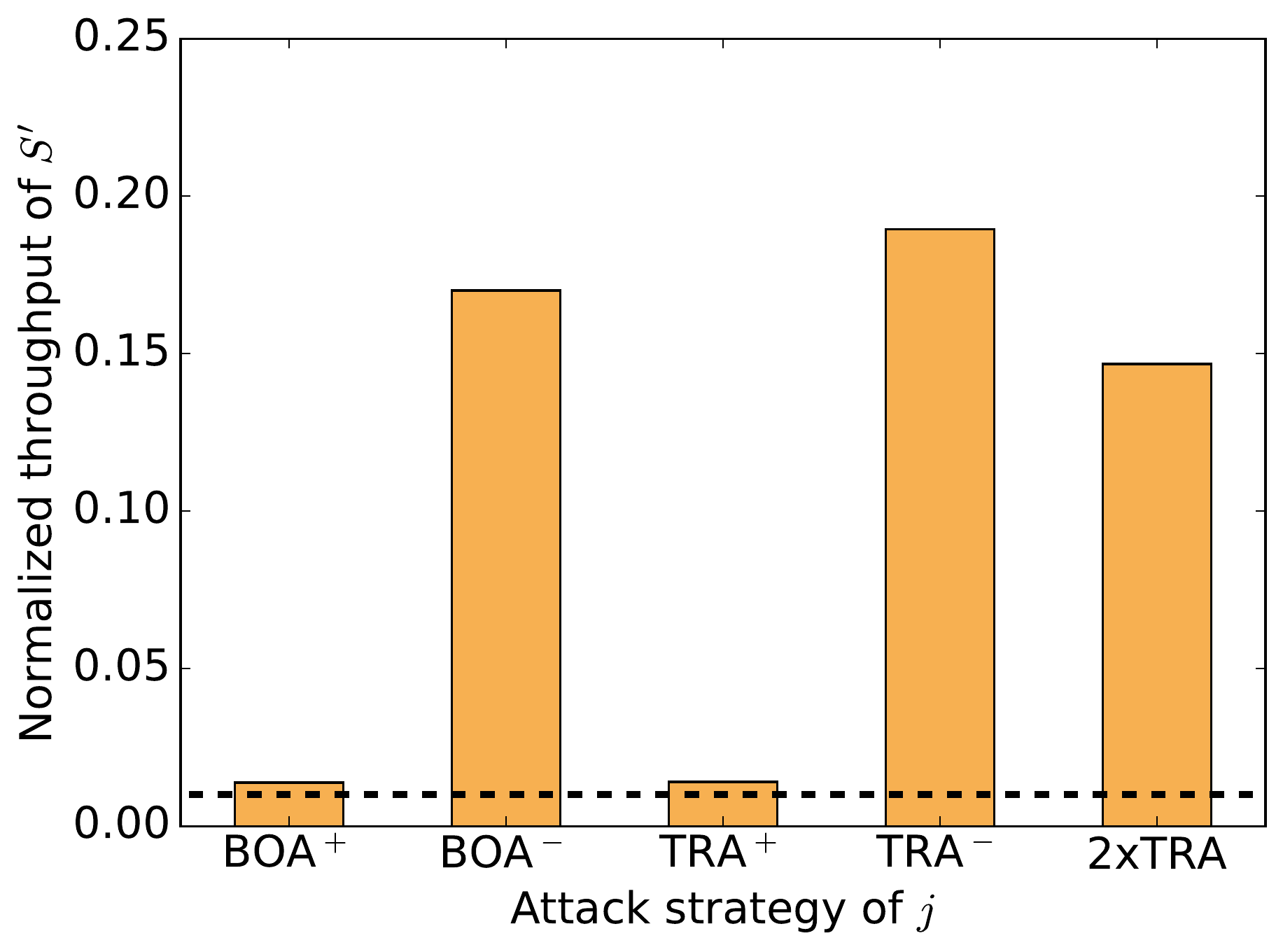}
\caption{}\label{fig:singlehop-sat-dl-j}
\end{subfigure}
\begin{subfigure}[]{0.32\textwidth}
\includegraphics[width=\textwidth]{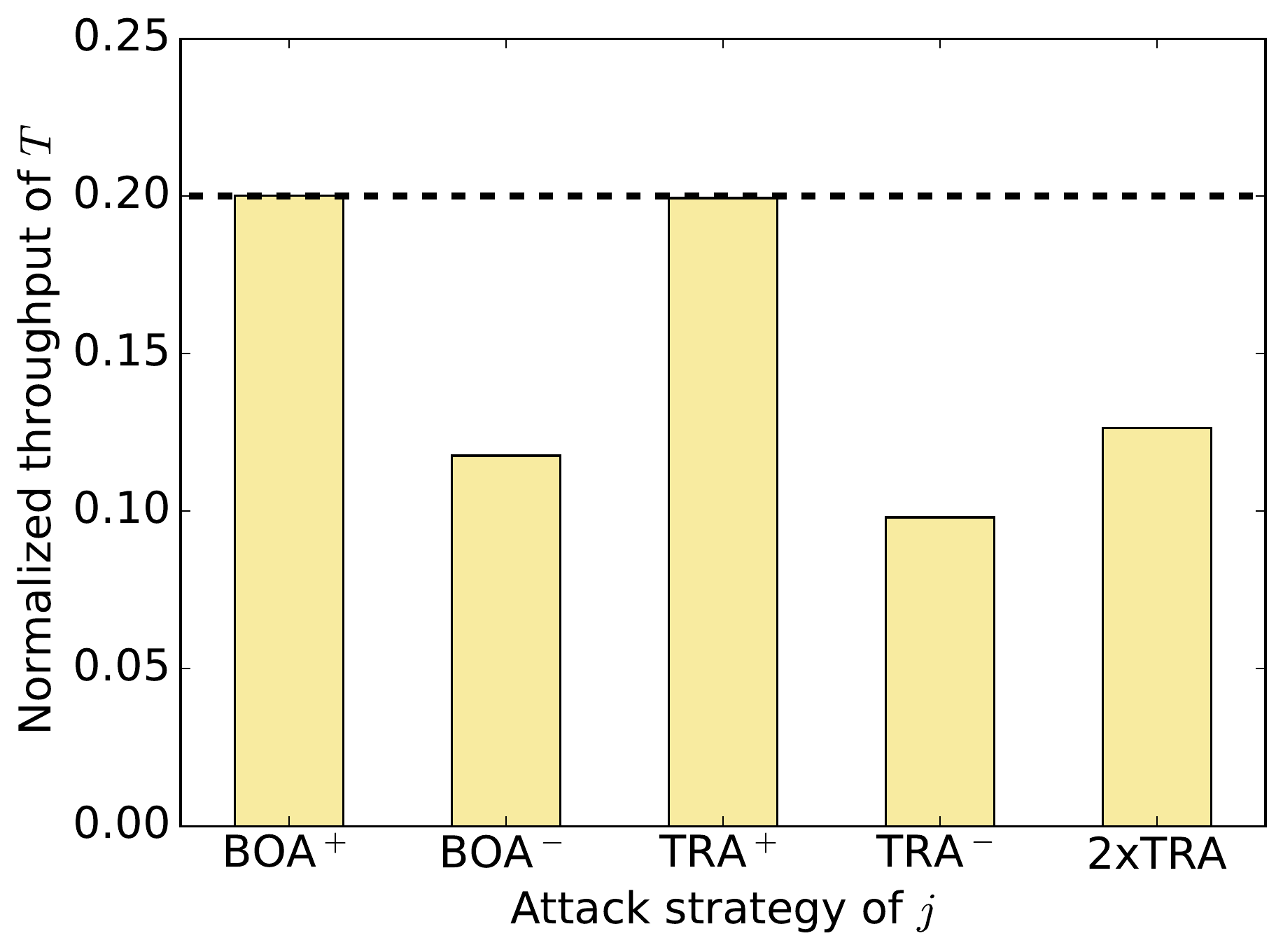}
\caption{}\label{fig:singlehop-sat-dl-i}
\end{subfigure}
\begin{subfigure}[]{0.32\textwidth}
\includegraphics[width=\textwidth]{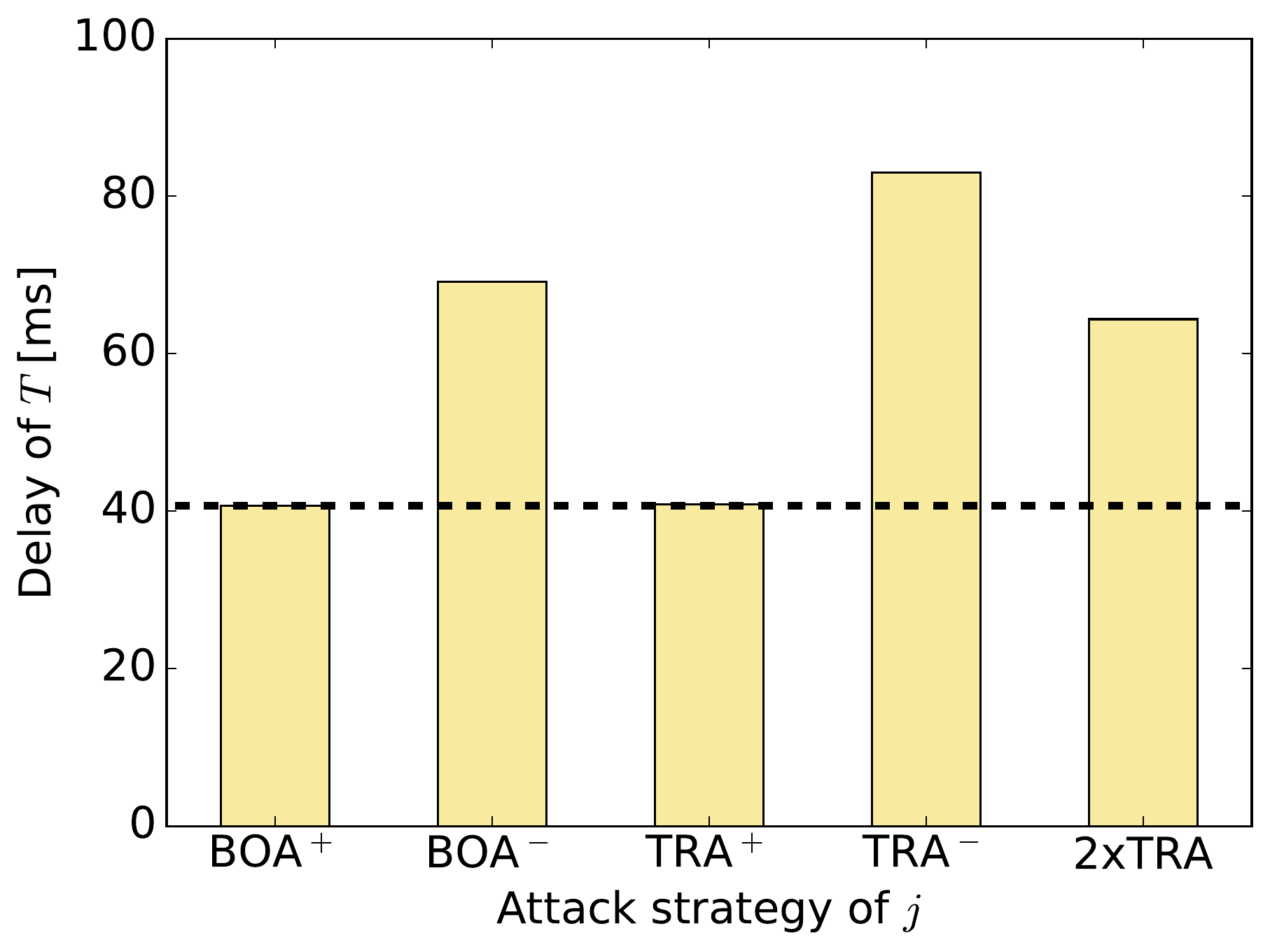}
\caption{}\label{fig:singlehop-sat-dl-del}
\end{subfigure}
\caption{Impact of A's attack strategy in the uplink (top) and downlink (bottom) scenario of the Fig.~\ref{fig:topology} topology: throughput of $S$/$S'$ (left), throughput of $T$ (middle), and delay of $T$ (right). Throughput is normalized to the PHY data rate. 95\% confidence intervals are either shown or too small for graphical representation. The reference line is for the case of A's honest behavior.}
\label{fig:singlehop-ul}
\vspace{-3ex}
\end{figure*}

In the uplink scenario, from A's perspective, significant throughput gains of $S$ can be achieved by attacking (Fig.~\ref{fig:singlehop-sat-ul-j}). However, it is visible that BOA$^+$ (where A uses both BE and VO queues configured with VO parameters) is not beneficial in two-hop relay settings. The reason is that, in addition to the inter-queue contention that worsens the overall performance, the VO queue used by $T$ retains its relative priority: the TCP ACKs for $T$ are handled as VO traffic at the AP, creating a smaller round-trip time than that experienced by $S$ (whose TCP ACKs are handled as BE traffic). Under TRA$^+$, flow $S$ approaches the throughput achieved by $T$ with A's honest behavior (the reference line in Fig.~\ref{fig:singlehop-sat-ul-i}). The downgrading attacks provide even better gains, which supports the hypothesis that upgrading source traffic is less important than downgrading transit traffic and elimination of inter-queue contention (note that compared to BOA$^-$, TRA$^-$ yields slightly better gains because it causes the AP to handle TCP ACKs for $T$ as BE traffic). However, 2xTRA is strikingly beneficial.

The throughput gains of $S$ are accompanied by the loss in throughput by $T$ (Fig.~\ref{fig:singlehop-sat-ul-i}). In most cases the loss ranges between 30\% and 50\% with the exception of 2xTRA, where 90\% of the throughput is lost. Delay in all but the last case is below the ITU-T requirement of 100 ms (Fig.~\ref{fig:singlehop-sat-ul-del}). This shows that some selfish attacks can be harmless (inflict no QoS degradation for $T$), even in saturation conditions.

\subsection{Downlink Scenario}

In the downlink scenario 
both BOA$^+$ and TRA$^+$ provide a small (3--5\%) throughput gain for $S'$ by increasing the sending rate of the corresponding TCP ACKs at the price of increased collision rate due to smaller CW values (Fig.~\ref{fig:singlehop-sat-dl-j}). BOA$^-$ and TRA$^-$ reduce the collision rate due to larger CW values, therefore are much more beneficial for A; BOA$^-$ slightly less so because, cf. Section~\ref{sec:uplink}, the VO queue used by $T$ still has relative priority despite being configured with BE parameters, whereas TRA$^-$ eliminates the inter-queue contention at A. Unexpectedly for A, 2xTRA performs no better than TRA$^-$. The reason is that the TRA$^+$ component has TCP ACKs for flow $S'$ handled as VO traffic at A. Thus $S'$ experiences a lower round-trip time and its transmit window is excessively expanded; the resulting increased collisions between AP and relay transmissions ultimately lower the throughput of $S'$.
In all considered downlink cases, $T$'s delay, though sometimes elevated beyond the reference value (representing A's honest behavior), meets the ITU-T requirement of 100 ms (Fig.~\ref{fig:singlehop-sat-dl-del}).

\section{Defense Measures}
\label{sec:countermeasures}

Once detected \cite{Patras2016, Konorski2014}, selfish MAC-layer attacks launched by the relay station A cannot be simply punished by banning the attacker from further communication (by deauthentication and blacklisting), as this would mean loss of network access for station B. Revoking A's privileged status at the AP is one possibility. However, a more convenient, self-regulatory approach would be to provide incentives for A's honest cooperation. For example, if the AP suspects an ongoing attack launched by A, it can drop ACK frames for its source packets or shape its source traffic; both these punishment measures can be considered a subtle form of denial of service. We evaluate them in the topology of Fig.~\ref{fig:topology} in the uplink scenario to show that they involve only a small computational overhead on behalf of the punisher AP and no transmission overhead at all.

The first measure, dropping ACK frames \cite{cagalj2005selfish}, has the punisher refrain from sending MAC-layer ACK frames for correctly received data frames but only those belonging to the attacker's flow ($S$ in the uplink scenario and $S'$ in the downlink scenario).
The degree of penalty can be scaled by acknowledging an $\alpha \in [0,1]$ portion of frames.

The second measure, traffic shaping, has the punisher apply traffic control to TCP flows related to the attacker's  traffic ($S$ in the uplink and $S'$ in the downlink scenario), e.g., in the form of a leaky bucket filter with a controlled output rate.
This rate can be proportional, by $\alpha \in [0,1]$, to the attacker's rate during an attack, so that for $\alpha=1$ the throughput of $S$ is equal to that shown in Fig.~\ref{fig:singlehop-sat-ul-j}.

\begin{figure}[tbp]
\centering

	\includegraphics[width=0.4\textwidth,trim=0 0 10ex 8ex, clip]{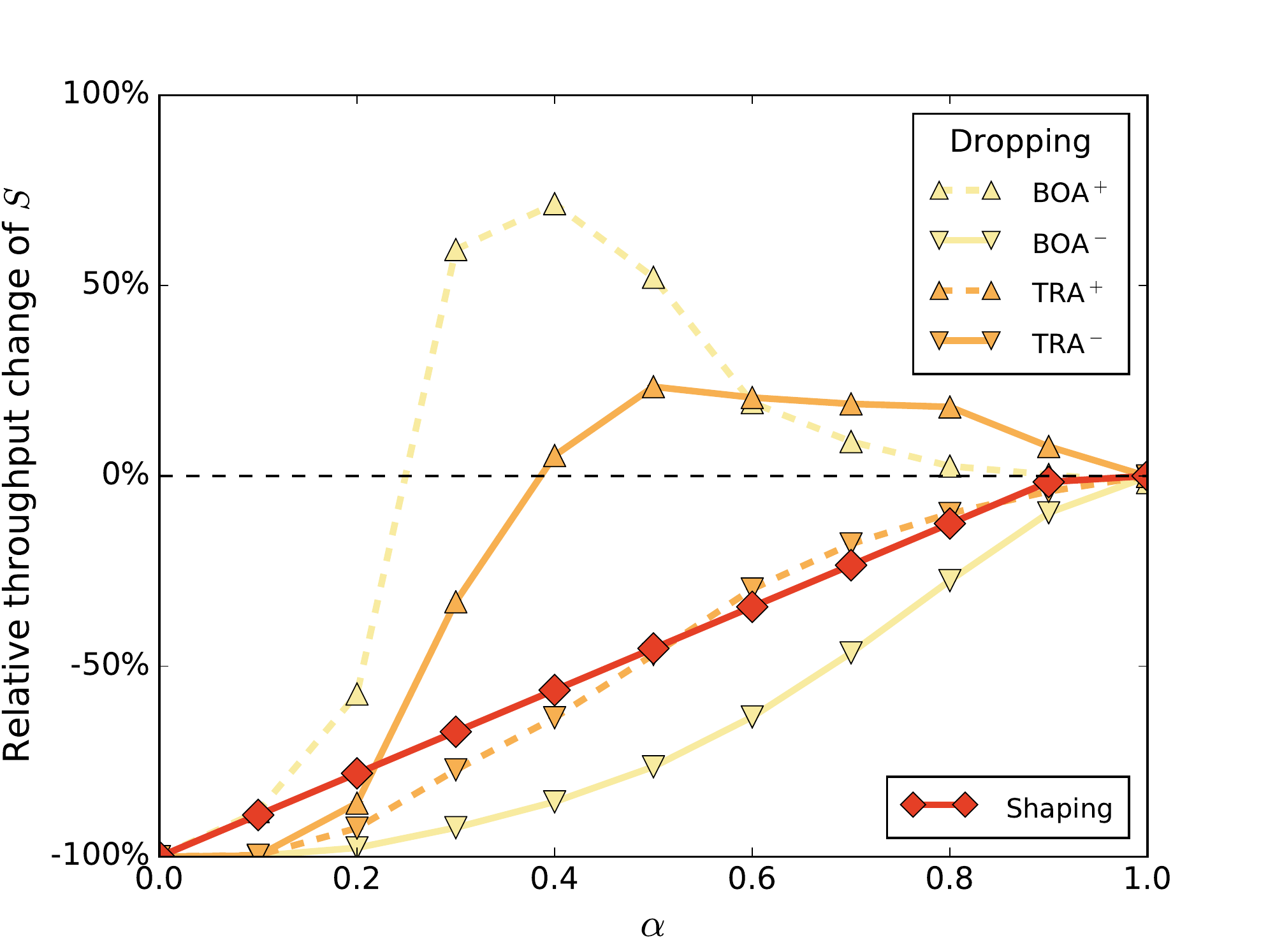}
\caption{Change of throughput of 
$S$ caused by punishment at the AP in comparison to unpunished attack.}
\label{fig:selective}
\end{figure}

The results presented in Fig.~\ref{fig:selective} show that
ACK dropping was able to scale $S$'s throughput for the downgrading attacks.
However, for the upgrading attacks, dropping ACKs can inadvertently optimize $S$'s TCP flow with respect to the effect of hidden stations\footnote{Careful pacing of TCP transmissions could improve flow S's performance in the presence of hidden stations even without ACK dropping, although it might necessitate cross-layer optimization \cite{Xie2015}.} 
and thus cause an unexpected \emph{increase} of $S$'s throughput for $\alpha \in [0.4,0.9]$.
We have also evaluated the performance of flow $T$ (not shown) and observed that for TRA$^-$ the throughput is reduced for both $S$ and $T$; clearly, ACK dropping has caused \textit{network} performance degradation.
In contrast, traffic shaping allowed to selectively (and almost linearly) control the throughput of $S$.
Similar results (not presented here) were obtained in the downlink scenarios.

\section{Conclusions}
\label{sec:conclusions}

We conclude the following: 
\begin{inparaenum}[1)]
\item except for 2xTRA in the uplink scenario, station A's selfish attacks were found harmless, since they did not violate the QoS requirements of high-priority traffic,
\item unlike in single-hop networks
BOA$^+$ brings the attacker no benefit in two-hop relay networks because it does not modify the packet's QoS designation,
\item downgrading attacks outperform their upgrading counterparts (particularly in the downlink scenario), while 
combined attack strategies were found to be only beneficial in one  case (2xTRA in uplink), 
\item simple ACK dropping 
is not a valid punishment in two-hop relay networks because of its unpredictable behavior; certainly, more sophisticated ACK dropping needs to be studied in the future. Future work should also consider more advanced incentive methods for relays, practical evaluation of the proposed defense methods both under TCP and UDP traffic, and analysis of new attack vectors enabled by 802.11ac/ax.
\end{inparaenum}

\vspace{-2ex}

%
%

\ifCLASSOPTIONcaptionsoff
  \newpage
\fi



\bibliographystyle{IEEEtran}

\begin{thebibliography}{1}
\providecommand{\url}[1]{#1}
\csname url@samestyle\endcsname
\providecommand{\newblock}{\relax}
\providecommand{\bibinfo}[2]{#2}
\providecommand{\BIBentrySTDinterwordspacing}{\spaceskip=0pt\relax}
\providecommand{\BIBentryALTinterwordstretchfactor}{4}
\providecommand{\BIBentryALTinterwordspacing}{\spaceskip=\fontdimen2\font plus
\BIBentryALTinterwordstretchfactor\fontdimen3\font minus
  \fontdimen4\font\relax}
\providecommand{\BIBforeignlanguage}[2]{{%
\expandafter\ifx\csname l@#1\endcsname\relax
\typeout{** WARNING: IEEEtran.bst: No hyphenation pattern has been}%
\typeout{** loaded for the language `#1'. Using the pattern for}%
\typeout{** the default language instead.}%
\else
\language=\csname l@#1\endcsname
\fi
#2}}
\providecommand{\BIBdecl}{\relax}
\BIBdecl

\bibitem{Garcia-Saavedra2015}
A.~Garcia-Saavedra, B.~Rengarajan, P.~Serrano, D.~Camps-Mur, and
  X.~Costa-Perez, ``{SOLOR}: Self-optimizing {WLANs} with legacy-compatible
  opportunistic relays,'' \emph{IEEE/ACM Transactions on Networking}, vol.~23,
  no.~4, pp. 1202--1215, 2015.

\bibitem{Konorski2014}
J.~Konorski and S.~Szott, ``Discouraging traffic remapping attacks in local ad
  hoc networks,'' \emph{IEEE Transactions on Wireless Communications}, vol.~13,
  pp. 3752--3767, 2014.

\bibitem{cagalj2005selfish}
M.~Cagalj, S.~Ganeriwal, I.~Aad, and J.-P. Hubaux, ``{On selfish behavior in
  CSMA/CA networks},'' in \emph{Proc. of INFOCOM}, 2005.

\bibitem{Patras2016}
P.~Patras, H.~Feghhi, D.~Malone, and D.~J. Leith, ``{Policing 802.11 MAC
  Misbehaviours},'' \emph{IEEE Transactions on Mobile Computing}, vol.~15,
  no.~7, pp. 1728--1742, 2016.

\bibitem{Konorski2017}
J.~Konorski and S.~Szott, ``Modeling a traffic remapping attack game in a
  multi-hop ad hoc network,'' in \emph{Proc. of IEEE GLOBECOM}, 2017.

\bibitem{Xie2015}
H.~Xie and A.~Boukerche, ``{TCP-CC: cross-layer TCP pacing protocol by
  contention control on wireless networks},'' \emph{Wireless Networks},
  vol.~21, no.~4, pp. 1061--1078, 2015.

\end{thebibliography}



%

%






\end{document}